\documentclass[paper a4]{article}
\title{Noncommutative version of an arbitrary nondegenerated mechanics.}
\author{A.A. Deriglazov\footnote{alexei@fisica.ufjf.br ~ On leave of
absence from Dept. Math. Phys., Tomsk Polytechnical University,
Tomsk, Russia.}}
\date{Dept. de Matematica, ICE, Universidade Federal de Juiz de Fora,\\
MG, Brasil.}
\begin{document}
\maketitle
\large
\begin{abstract}
A procedure to obtain noncommutative version for any nondegenerated
dynamical system is proposed and discussed.
\end{abstract}


\noindent
It is known that the
noncommutative geometry [1, 2] of the position variables in some mechanical
models can be obtained [3-8] as the result of direct canonical quantization
[9, 10] of underlying dynamical
systems with second class constraints. Nontrivial bracket for
the position variables appears in this case as the Dirac bracket,
after taking into account the constraints presented in the model.
In this note we show how to
obtain the noncommutative version of any nondegenerated mechanical system.
Namely, the following statement will be demonstrated.

Let $S=\int dt L(q^A, ~ \dot q^A)$ is action of some nondegenerated system,
and let $L_1(q^A, ~ \dot q^A, ~ v_A)$ is
the corresponding first order Lagrangian (see below). Then the corresponding
noncommutative version is
$S_N=\int dt\left[ L_1(q^A, ~ \dot q^A, ~  v_A)+
\dot v_A\theta^{AB}v_B\right]$.
Namely, the system $S_N$ has the following properties: \\
1) It has the same number of physical degrees of freedom as the initial
system $S$. \\
2) Equations of motion of the system are the same as for the initial
system $S$, modulo the term which is proportional to the parameter
$\theta^{AB}$. \\
3) Configuration space variables have the noncommutative brackets:
$\{q^A, ~ q^B\}=-2\theta^{AB}$.

We point also that quantization of the system $S_N$ leads to quantum
mechanics with ordinary product replaced by the Moyal product,
similarly to the case of a particle on noncommutative plane.

Let us present details. Our starting point is some nondegenerated
mechanical system with the configuration space variables $q^A(t), ~
A=1,2, \ldots , n$, and the Lagrangian action
\begin{eqnarray}\label{1}
S=\int dt L(q^A, ~ \dot q^A).
\end{eqnarray}
Due to nondegenerate character of the system, there are no constraints
in the Hamiltonian formulation. Let $p_A$ are conjugated momentum for
$q^A$, one can write the Hamiltonian action
\begin{eqnarray}\label{2}
S_H=\int dt \left[ p_A\dot q^A-H_0(q^A, ~ p_A)\right].
\end{eqnarray}
Equations of motion which follow from Eq.(\ref{1}) and (\ref{2}) are
equivalent (they remain equivalent for any degenerated system also
[10, 11]). Equivalently,
one can describe the initial system (\ref{1}) by means of the first order
Lagrangian action
\begin{eqnarray}\label{3}
S_1=\int dt \left[v_A\dot q^A-H_0(q^A, ~ v_A)\right].
\end{eqnarray}
Here $q^A(t), ~ v_A(t)$ are the configuration space variables of the
formulation \footnote{The Lagrangian formulations (\ref{1}), (\ref{3})
are equivalent. Actually, denoting the conjugated momentum for the
variables $q^A, ~ v_A$ as $p_A, ~ \pi^A$ one finds, in the Hamiltonian
formulation for the action (\ref{3}), the second class constraints
$p_A-v_A=0, ~ \pi^A=0$. Introducing the corresponding Dirac bracket,
one can treat the constraints as the strong equations. Then the
Hamiltonian formulation for (\ref{3}) is the same as for (\ref{1}),
namely Eq.(\ref{2}).}.
The
noncommutative version of the system (\ref{1}) is described by the
following Lagrangian action
\begin{eqnarray}\label{4}
S_N=\int dt \left[v_A\dot q^A-H_0(q^A, ~ v_A)+
\dot v_A\theta^{AB}v_B \right],
\end{eqnarray}
where $\theta^{AB}$ is some constant matrix. It turns out to be the
noncommutativity parameter for the variables $q^A$.

To analyse the physical sector of the Lagrangian system (\ref{4}),
we rewrite it in the Hamiltonian form. All the expressions for determining
of the momentum turn out to be the primary constraints of the model ~
($p_A,  \pi^A$ are conjugated momentum for the variables $q^A,  v_A$)
\begin{eqnarray}\label{5}
G_A\equiv p_A-v_A=0, \qquad
T^A\equiv\pi^A-\theta^{AB}v_B.
\end{eqnarray}
The Hamiltonian is
\begin{eqnarray}\label{6}
H=H_0(q^A, ~ v_A)+\lambda_{1}^AG_A+\lambda_{2A}T^A,
\end{eqnarray}
where $\lambda$ are the
Lagrangian multipliers for the constraints. On the next step of the
procedure there are appear only equations for determining of the Lagrangian
multipliers
\begin{eqnarray}\label{7}
\lambda_{2A}=-\frac{\partial H_0}{\partial q^A}, \qquad
\lambda_1^A=\frac{\partial H_0}{\partial v_A}-2\theta^{AB}
\frac{\partial H_0}{\partial q^B}.
\end{eqnarray}
Equations of motion follow from (\ref{6}), (\ref{7})
\begin{eqnarray}\label{8}
\dot q^A=\frac{\partial H_0}{\partial v_A}-2\theta^{AB}
\frac{\partial H_0}{\partial q^B}, \quad
\qquad \dot p_A=-\frac{\partial H_0}{\partial q^A}; \cr
\dot v_A=-\frac{\partial H_0}{\partial q^A}, \quad \quad
\qquad \dot \pi^A=-\theta^{AB}\frac{\partial H_0}{\partial q^B}.
\end{eqnarray}
They are accompanied by the second class constraints (\ref{5}).
Poisson brackets of the constraints are
\begin{eqnarray}\label{9}
\{G_A, G_B\}=0, \qquad \{T^A, T^B\}=
-2\theta^{AB}, \qquad
\{G_A, T^B\}=-\delta_A^B.
\end{eqnarray}
The constraints can be taken into account by transition to the Dirac
bracket. Introducing the Dirac bracket
\begin{eqnarray}\label{10}
\{A, B\}_D=\{A, B\}+2\{A, G_A\}\theta^{AB}
\{G_B, B\}- \cr
\{A, G_A\}\{T^A, B\}+\{A, T^A\}\{G_A, B\},
\end{eqnarray}
one finds, in particular, the following brackets for the
fundamental variables (all the nonzero brackets are presented)
\begin{eqnarray}\label{11}
\{q^A, q^B\}=-2\theta^{AB}, \qquad
\{q^A, p_B\}=\delta^A_B,
\qquad \{p_A, p_B\}=0;
\end{eqnarray}
\begin{eqnarray}\label{12}
\{q^A, v_B\}=\delta^A_B, \quad
\{q^A, \pi^B\}=-\theta^{AB}.
\end{eqnarray}
One has now different possibilities to choose the physical sector: either
$(q^A, ~ p_A)$, or $(q^A, ~ v_A)$, or $(q^A, ~ \pi_A)$ ~
(the latter possibility implies that $\theta$ is invertible).
Let us take the variables $(q^A, ~ p_A)$ ~ (the same as for the initial
formulation (\ref{1})) as the physical one. Since the
Dirac bracket has been introduced,
the variables $v, \pi$ can be omitted from consideration.
Dynamics of the physical variables is governed now by the equations
\begin{eqnarray}\label{13}
\dot q^A=\frac{\partial H_0}{\partial p_A}-2\theta^{AB}
\frac{\partial H_0}{\partial q^B}, \qquad
\qquad \dot p_A=-\frac{\partial H_0}{\partial q^A},
\end{eqnarray}
where $H_0(q, ~ p)=H_0(q, ~ v)|_{v\rightarrow p}$.
Modulo the term with $\theta$, they are the same as for the initial
system (\ref{1}). Brackets for the variables $q^A, ~ p_A$ are given
by Eqs.(\ref{11}). One can show that other possibilities to
choose the physical variables lead to an equivalent description.

To quantize the resulting system, one possibility is to find variables
which have the standard brackets. For the case under consideration
they are
\begin{eqnarray}\label{14}
\tilde q^A=q^A-\theta^{AB}p_B, \qquad \tilde p_A=p_A,
\end{eqnarray}
and obey $\{\tilde q, \tilde q\}=\{\tilde p, \tilde p\}=0, ~
\{\tilde q, \tilde p\}=1$. Equations of motion in terms
of these variables acquire the standard form
\begin{eqnarray}\label{15}
\dot{\tilde q}^A=\{\tilde q^A, \tilde H_0\}, \qquad
\dot{\tilde p}_A=\{\tilde p_A, \tilde H_0\},
\end{eqnarray}
where $\tilde H_0=H_0(\tilde q+\theta\tilde p, ~ \tilde p)$. It leads to
quantum
mechanics with the Moyal product (see [7] and references therein)
\begin{eqnarray}\label{16}
H_0(\tilde q^A+\theta^{AB}\tilde p_B, ~ \tilde p_B)\Psi(\tilde q^C)=
H_0(\tilde q^A, ~ \tilde p_B)*\Psi(\tilde q^C).
\end{eqnarray}

In conclusion, let us point that the procedure described above can be
applied to some degenerated systems as well. In particular, the
noncommutative relativistic particle has been proposed in [8]
following a similar line.

\end{document}